\documentclass[%
 aip,
 amsmath,amssymb,
 reprint,%
]{revtex4-1}

\def\ispreprint{1} 

\usepackage{graphicx}
\usepackage{dcolumn}
\usepackage{placeins}

\usepackage[utf8]{inputenc}
\usepackage[T1]{fontenc}
\usepackage{mathptmx}
\usepackage{etoolbox}
\usepackage{xcolor}
\usepackage{SIunits}
\usepackage[textsize=tiny]{todonotes}
\usepackage{bbm}

\newcommand{\paragraphtitle}[1]{{\small\textsf{\textbf{#1}}} \normalfont}

\usepackage{changes} 

\makeatletter
\def\@email#1#2{%
 \endgroup
 \patchcmd{\titleblock@produce}
  {\frontmatter@RRAPformat}
  {\frontmatter@RRAPformat{\produce@RRAP{*#1\href{mailto:#2}{#2}}}\frontmatter@RRAPformat}
  {}{}
}%
\makeatother
\begin{document}

\preprint{AIP/123-QED}

\title{Combining transition path sampling with data-driven collective variables through a reactivity-biased shooting algorithm}

\author{Jintu Zhang}
\affiliation{Innovation Institute for Artificial Intelligence in Medicine of Zhejiang University, College of Pharmaceutical Sciences, Zhejiang University, Hangzhou 310058 Zhejiang, China}
\affiliation{Atomistic Simulations, Italian Institute of Technology, Genova 16152, Italy}

\author{Odin Zhang}
\affiliation{Innovation Institute for Artificial Intelligence in Medicine of Zhejiang University, College of Pharmaceutical Sciences, Zhejiang University, Hangzhou 310058 Zhejiang, China}

\author{Luigi Bonati$^*$}
\email{luigi.bonati@iit.it}
\affiliation{Atomistic Simulations, Italian Institute of Technology, Genova 16152, Italy}

\author{TingJun Hou$^*$}
\email{tingjunhou@zju.edu.cn}
\affiliation{Innovation Institute for Artificial Intelligence in Medicine of Zhejiang University, College of Pharmaceutical Sciences, Zhejiang University, Hangzhou 310058 Zhejiang, China}
\affiliation{State Key Lab of CAD\&CG, Zhejiang University, Hangzhou, Zhejiang 310058, China}


\begin{abstract}
Rare event sampling is a central problem in modern computational chemistry research. Among the existing methods, transition path sampling (TPS) can generate unbiased representations of reaction processes. However, its efficiency depends on the ability to generate reactive trial paths, which in turn depends on the quality of the shooting algorithm used. We propose a new algorithm based on the shooting success rate, \textit{i.e.} reactivity,  measured as a function of a reduced set of collective variables (CVs). These variables are extracted with a machine learning approach directly from TPS simulations, using a multi-task objective function. Iteratively, this workflow significantly improves shooting efficiency without any prior knowledge of the process. In addition, the optimized CVs can be used with biased enhanced sampling methodologies to accurately reconstruct the free energy profiles. We tested the method on three different systems: a two-dimensional toy model, conformational transitions of alanine dipeptide, and hydrolysis of acetyl chloride in bulk water.
In the latter, we integrated our workflow with an active learning scheme to learn a reactive machine learning-based potential, which allowed us to study the mechanism and free energy profile with an \textit{ab initio}-like accuracy.
\end{abstract}

\maketitle
\if\ispreprint1

\begin{figure}[b!]\centering
\includegraphics{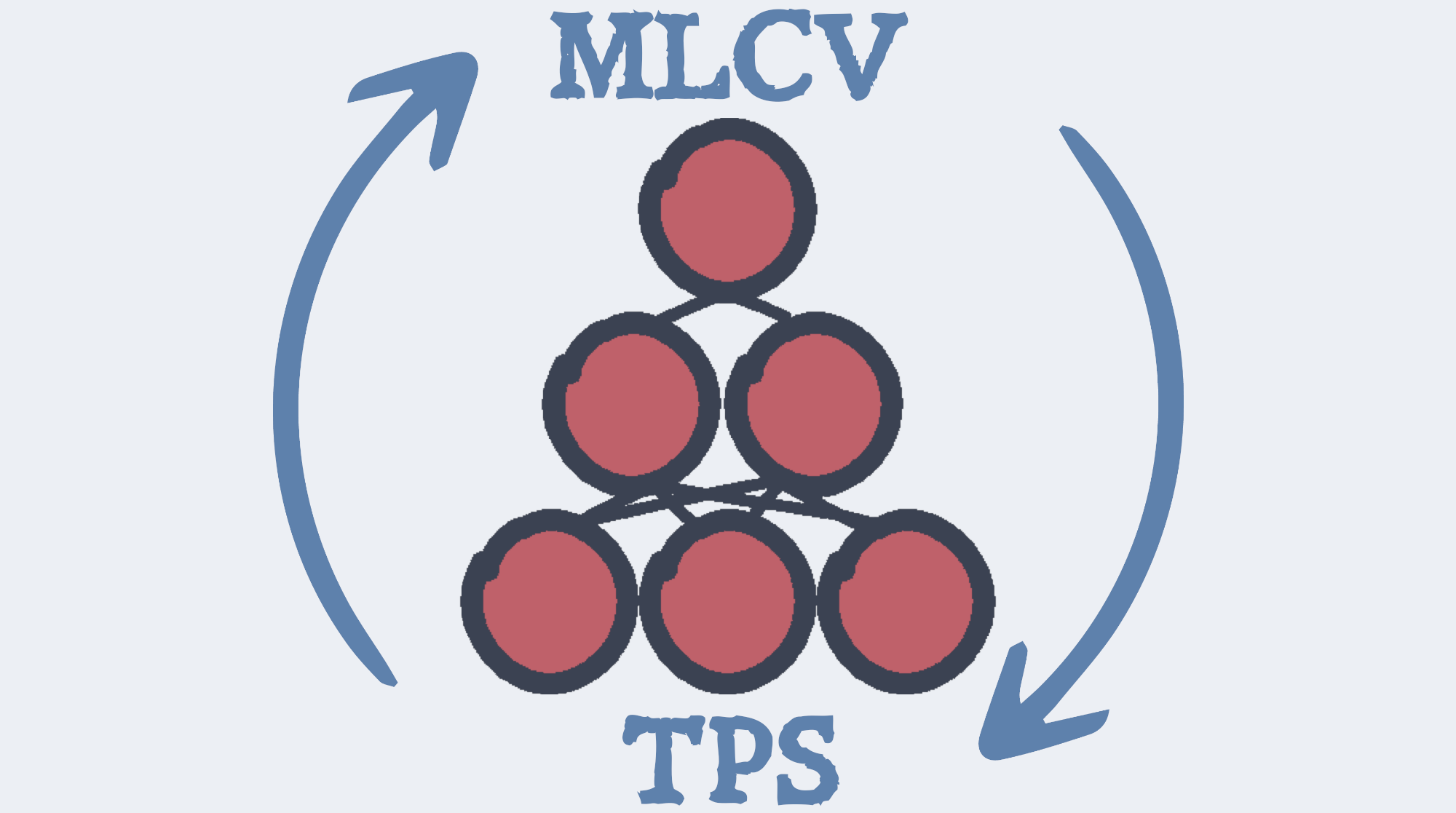}
\end{figure}
\else

\fi

\section{Introduction}
Molecular dynamics (MD) has become a standard protocol in the study of dynamical processes in physics, chemistry, biology and pharmaceutical science~\cite{frenkel2023understanding}.
However, many relevant processes are related to transitions between dynamically distinct phase space regions (``states'').
Typical relaxation times may range from several nanoseconds to tens of seconds, according to the underlying mechanisms.
Therefore, these conformational transitions are usually referred to as ``rare events''.
While in principle very long MD simulations may provide both stationary and dynamical properties\cite{shaw2009,shaw2014}, ergodic trajectories are not generally available. This is even more challenging when the computational scaling of potential energy evaluations is extremely high \cite{parrinello1985}.
For all these reasons, rare events bring significant challenges to the study of atomistic systems via MD simulations.

To address this problem, in the last decades, various enhanced sampling methods have been presented \cite{schulten2015,gao2019,bolhuis2009,bolhuis2015}.
They can be roughly divided into two categories, depending on whether they artificially modify the system's potential energy surface.
The first category includes biased simulation protocols, e.g., umbrella sampling (US) \cite{torrie1977}, Metadynamics (MetaD) \cite{parrinello2002}, on-the-fly probability enhanced sampling (OPES) \cite{parrinello2020}, Gaussian accelerated molecular dynamics (GaMD) \cite{miao2014}, etc.
These methods can effectively generate ergodic trajectories, and using suitable estimators \cite{parrinello2015,chodera2008,miao2014}, it is possible to reconstruct the free energy profiles of the system.
Still, biased simulation methods perturb the dynamics of the simulated system.
Although the kinetics of the processes of interest can be restored by elaborate reweighting schemes \cite{parrinello20231}, mechanisms discovered by biased simulations may still diverge from the true ones, especially when the biasing potentials are applied to suboptimal reaction coordinates (RC).
The other family of methods focuses on sampling paths between different states and is commonly known as ``path sampling'' methods. These approaches can preserve the dynamics of the simulated system and generate more natural trajectories.
Thus, they are more suitable for investigating the mechanism and for reaction rate calculations.
Notable examples are transition path sampling (TPS) \cite{bolhuis1998}, forward flux sampling (FFS) \cite{allen2005}, weighted ensemble (WE) \cite{huber1996} and milestoning \cite{fadadjian2004}. In addition, some approaches combine aspects of the two families, such as collective path variables~\cite{Branduardi2007,Ortiz2018}. These variables are built to describe the pathways connecting the states, but they are typically used with biased sampling algorithms. Recently, machine learning-based approaches have also been used to construct them in a data-driven manner~\cite{Pietrucci2023,frohlking2024deep}.

Within path sampling approaches, TPS performs Markov Chain Monte Carlo (MCMC) in the path space to provide an unbiased ensemble of reactive trajectories\cite{bolhuis2021}.
Since the generated paths do not contain relaxations within the ending-point states, it is called Transition Path Ensemble (TPE).
In principle, 
the only requirements for setting up a TPS simulation are an initial reactive path and coordinates that distinguish different ending-point states.
Unlike biased sampling methods, TPS does not require the chosen coordinates to reflect the true ``slow modes'' of the systems since the MCMC sampling will ensure the correct statistics of the TPE.
However, as with any other Monte Carlo method, TPS requires algorithms that generate trial samples from old samples, known as MC moves.
For TPS, one of the most effective and widely used is the shooting move \cite{bolhuis19981}.
This move selects a snapshot (the shooting point) from the old path and perturbs it with a tractable probability, which will be used to calculate the acceptance criterion of this move.
After the shooting point perturbation, a trial path is created from the modified snapshot by propagating the system forward and backward in time until a metastable state is reached.
If the newly obtained path is not reactive, it will be discarded, and another trial path will be generated.
The efficiency of TPS simulations depends on the ability of the chosen shooting strategy to generate reactive paths.
The ordinary two-way shooting move could be inefficient when the dynamics is less deterministic or the reactive paths are very long.
Under these situations, shooting points far from the transition region are not likely to lead to reactive paths since the effect of initial velocities will soon be lost during the stochastic trajectory propagation.
Thus, selecting shooting points from the transition region seems to be a natural strategy to improve the efficiency of TPS.
As shown by Jung \textit{et. al.}~\cite{jung2017}, the ``shooting from the top'' scheme significantly improves the efficiency of reactive trajectory generation.
However, in the same work, it has been pointed out that misplacing the shooting point selection range (from now on referred to as ``the shooting range'') may harm the sampling efficiency.
That is to say, to benefit from this sampling scheme, not only should the chosen reaction coordinate be able to distinguish the transition state from the metastable ones, but also, the location of the transition region on the reaction coordinate should be known in advance. These requirements are obviously in conflict with the spirit of TPS.
To resolve these difficulties, Jung and Covino \textit{et. al.} introduced the Artificial Intelligence for Molecular Mechanism Discovery (AIMMD) workflow~\cite{jung20231,jung20232}. 
In AIMMD, the reaction coordinate and the shooting range are iteratively optimized based on the committor function\cite{bolhuis2000,hummer2005}.
This is learned during the optimization in a data-driven way, similar to the 
work of Ref.\cite{bolhuis20211}, where transition paths sampled by TPS were used to learn committor functions via an extended autoencoder.
Furthermore, in AIMMD, the committor probabilities are used to bias the shooting ranges via a symmetric smearing function centered at the value of 0.5, associated with the transition state~\cite{jung20231}.
Consequently, this method can generate shooting points close to the transition region, thereby markedly enhancing the efficiency of TPS simulations. Indeed, even for very complex systems, AIMMD achieves a high success rate in shooting moves~\cite{jung20231}. Furthermore, with an appropriate reweighting of the sampled TPE, AIMMD could also provide free energy profiles of the sampled reaction channels~\cite{jung20232}.

In this study, we propose an alternative shooting move, which we call ``reactivity-biased shooting'' (RB-shooting), to promote the effectiveness of TPS simulations.
Our workflow is also based on the concept of shooting from the transition region, but what changes is how the area is identified. Instead of approximating the committor probability, we fit the density of shooting points in a low-dimensional space identified by one or more collective variables (CVs). The reactivity of a given shooting point, defined as the ratio of successful to total shooting point densities, is then iteratively used to influence the shooting range.
To reduce the dependence on prior knowledge about the simulated process, we used machine learning methods\cite{parrinello20232} to extract the CVs directly from simulation data. 
These variables are trained with a semi-supervised approach to classify the initial and end states and perform dimensionality reduction for all other points in the transition path ensemble. Although the criteria used to find CVs are simple and do not attempt to optimize the committor probability directly, this data-driven strategy proves very effective when supplied with the relevant data on reactive transitions.
Furthermore, the extracted CVs can be directly applied to perform enhanced sampling simulations, e.g., OPES or umbrella sampling, which allow us to obtain free energy profiles of the sampled reaction channels.
To illustrate the performance of our workflow, we carried out RB-shooting TPS simulations to study the motion of a Brownian particle on a 2-dimensional potential energy surface, conformational transitions of alanine dipeptide and the hydrolysis of acetyl chloride in explicit waters.
For each process, our workflow achieved very high shooting efficiency.
Besides, we also present the combination between our workflow and the fitting process of machine learning potentials when an \textit{ab initio} accuracy is required.

The paper is organized as follows.
In the methods section, we first introduce the reactivity-biased shooting move and then describe how we learn the collective variables from the TPS data. We then summarize the complete workflow in detail.
Subsequently, we describe the computational setup employed for the three systems and discuss the results obtained on each of them.
Finally, we outline some future perspectives.

\section{Methods}

\subsection*{Reactivity-Biased Shooting Point Selection.}

In MD simulations, a phase space path, or trajectory of length $L$, can be represented by a sequence of phase points: $\mathbf{x} = \{ x_0, x_1, ..., x_L \}$.
Given two disjoint phase space regions (states) $A$ and $B$, we consider a path as reactive if it starts in one state and ends in the other, \textit{i.e.} if it satisfies:

\if\ispreprint1
    \begin{align}
    \begin{split}
        \mathbbm{1}_{A}(x_0)\mathbbm{1}_{B}(x_L) &+ \mathbbm{1}_{A}(x_L)\mathbbm{1}_{B}(x_0) > 0 \\ \text{ and }& \\ \mathbbm{1}_{A}(x_i)\mathbbm{1}_{B}(x_i) &= 0\ \forall i \in [1, L - 1], 
    \end{split}
    \end{align}
    \label{eqn:1}
\else
    \begin{equation}
    \mathbbm{1}_{A}(x_0)\mathbbm{1}_{B}(x_L) + \mathbbm{1}_{A}(x_L)\mathbbm{1}_{B}(x_0) > 0 \text{ and } \mathbbm{1}_{A}(x_i)\mathbbm{1}_{B}(x_i) = 0\ \forall i \in [1, L - 1], \label{eqn:1}
    \end{equation}
\fi

where $\mathbbm{1}_{S}[x]$ is the indicator function for state $S$.
Under the assumption of Markovian dynamics, the equilibrium ensemble constructed by reactive paths (TPE) may be sampled by TPS.
To reproduce the correct distributions, TPS generates a sequence of paths by proposing trial samples and accepting them based on the Metropolis criterion.

In TPS, trial paths are generated by shooting moves.
The shooting move selects a point from the old path and modifies it to generate a new shooting point.
A common way is to change only the velocities of the shooting point while conserving its total kinetic energy (and total momentum) and keeping its configuration unchanged.
Thus, the shooting point's phase space density does not change during the move.
Then, a trial path is created from the new shooting point by propagating the system forward and backward in time until a metastable state is reached.
If the newly obtained path fulfills Eq.~1, it will be accepted or rejected according to the following criterion\cite{jung2017}:
\if\ispreprint1
    \begin{widetext}
    \begin{equation}
p_{acc}\big(\mathbf{x}^{(o)} \to \mathbf{x}^{(n)}\big) = \big(\mathbbm{1}_{A}(x_0)\mathbbm{1}_{B}(x_L) + \mathbbm{1}_{A}(x_L)\mathbbm{1}_{B}(x_0)\big) \min \Bigg[ \frac{p_{sel}\big(x_{sp};\ \mathbf{x}^{(n)}\big)}{p_{sel}\big(x_{sp};\ \mathbf{x}^{(o)}\big)}, 1 \Bigg]. \label{eqn:2}
\end{equation}
    \end{widetext}
\else
\begin{equation}
p_{acc}\big(\mathbf{x}^{(o)} \to \mathbf{x}^{(n)}\big) = \big(\mathbbm{1}_{A}(x_0)\mathbbm{1}_{B}(x_L) + \mathbbm{1}_{A}(x_L)\mathbbm{1}_{B}(x_0)\big) \min \Bigg[ \frac{p_{sel}\big(x_{sp};\ \mathbf{x}^{(n)}\big)}{p_{sel}\big(x_{sp};\ \mathbf{x}^{(o)}\big)}, 1 \Bigg]. \label{eqn:2}
\end{equation}
\fi

In the above equation $\mathbf{x}^{(n)}$ and $\mathbf{x}^{(o)}$ stand for the newly generated trial path and the original path, while $p_{sel}\ extensive (x_{sp};\ \mathbf{x}\big)$ is the probability of selecting the shooting point $x_{sp}$ out of the path $\mathbf{x}$. 
In practice, this selection probability can be drawn from an arbitrary biasing function that takes the phase space's geometry or reaction coordinate values as input.
For example, in the original shooting from the top scheme, the biasing function is the reciprocal of the number of frames the trajectory spends in the shooting range \cite{bolhuis2021}.

In RB-shooting, we want to bias this shooting point selection based on the reactivity of the different points. Given a shooting point configuration x, we define its reactivity as:
\begin{equation}
\mathcal{R}_x = \lim_{N^{path}_{total} \to \infty} \frac{N^{path}_{reactive}}{N^{path}_{total}}, \label{eqn:3}
\end{equation}
where $N^{path}_{reactive}$ and $N^{path}_{total}$ are the number of reactive paths and all trial paths shot from the configuration during infinite long TPS runs.
According to the definition of transition state \cite{hummer2004}, configurational space points with the highest reactivities correspond to the transition state ensemble, characterized by a committor value of 0.5.
This definition could be cast into the following form, equivalently:
\begin{equation}
\mathcal{R}_x = \lim_{N^{MC}_{total} \to \infty} \frac{\rho^{sp}_{reactive}(x)N^{MC}_{reactive}}{\rho^{sp}_{total}(x)N^{MC}_{total}}, \label{eqn:4}
\end{equation}
where $N^{MC}_{reactive}$ and $N^{MC}_{total}$ are the number of reactive MC moves and all MC moves, while $\rho^{sp}_{reactive}(x)$ and $\rho^{sp}_{total}(x)$ are configurational space densities of reactive and all shooting points at the required configuration $x$.
To achieve high shooting efficiency, it is natural to select the shooting points that bear high reactivities.
Thus, we introduce the following normalized biasing function to perform shooting point selection:
\begin{equation}
p_{sel}\big(x_i;\ \mathbf{x}\big) = \frac{\mathcal{R}_{x_i}}{\sum_{x_j \in \mathrm{x}}\mathcal{R}_{x_j}}. \label{eqn:5}
\end{equation}
Obviously, due to the finiteness of realistic simulations, the direct application of Eq.~\ref{eqn:3} or Eq.~\ref{eqn:4} is impossible.
Therefore, we make two approximations.
In the first one, we use the estimated shooting point densities $\tilde{\rho}^{sp}_{reactive}$ and $\tilde{\rho}^{sp}_{total}$ from a finite TPS run instead of the true densities.
Even so, the density fitting of shooting points in the high-dimensional configuration space could be extremely difficult.
We thus introduce the second approximation, which is to perform density estimation of the shooting points in the space identified by a set of low-dimensional reaction coordinates $\mathbf{s}(x)$. 

Based on the above approximations, we rewrite Eq.~\ref{eqn:5} as:
\begin{equation}
p_{sel}\big(x_i;\ \mathbf{x}\big) = \frac{1}{\mathcal{Z}}\frac{\tilde{\rho}^{sp}_{reactive}(\mathbf{s}(x_i))}{\tilde{\rho}^{sp}_{total}(\mathbf{s}(x_i))},
\ \mathcal{Z} = \sum_{x_j \in \mathrm{x}}\frac{\tilde{\rho}^{sp}_{reactive}(\mathbf{s}(x_j))}{\tilde{\rho}^{sp}_{total}(\mathbf{s}(x_j))}. \label{eqn:7}
\end{equation}
In practice, a kernel density estimate (KDE) is used to represent the densities, with the bandwidth parameter determined by Silverman’s rule \cite{silverman1998}. Furthermore, to reduce the noise caused by the density fitting in poorly sampled regions of the CVs space, we truncate the shooting point densities to construct the biasing function:

\begin{equation}
p_{sel}\big(x_i;\ \mathbf{x}\big) = \frac{1}{\mathcal{Z}}\frac{\hat{\rho}^{sp}_{reactive}(\mathbf{s}(x_i))}{\hat{\rho}^{sp}_{total}(\mathbf{s}(x_i))}
\label{eqn:8}
\end{equation}
with $\mathcal{Z} = \sum_{x_j \in \mathrm{x}}\frac{\hat{\rho}^{sp}_{reactive}(\mathbf{s}(x_j))}{\hat{\rho}^{sp}_{total}(\mathbf{s}(x_j))}$ and $\hat{\rho} =
\begin{cases}
  \tilde{\rho}, & \text{if}\ \tilde{\rho} > \text{cutoff} \\
  0, & \text{otherwise}
\end{cases}$.

Besides eliminating noises, increasing the cutoff value is identical to shrinking the size of the shooting range, thus improving the success rate of generating reactive paths.

To fulfill the second approximation, the chosen reaction coordinates should be able to identify the transition state structures from the reaction path. 
However, in the worst-case scenario where the chosen variables cannot capture the conformational transitions, the RB-shooting degenerates into standard two-way shooting.

\subsection*{Machine-learning collective variables from TPS data.}

Identifying appropriate collective variables is a critical issue in many enhanced sampling methods. Since constructing them based on physical intuition alone might be challenging, we build upon the recent developments in the machine learning collective variables (MLCV) \cite{parrinello20232} design.
In these approaches, one tries to learn directly from the data what degrees of freedom are essential, exploiting both unsupervised and supervised learning techniques.
Because an important requirement for the CVs is to distinguish the metastable states, numerous methods have addressed the problem of identifying them using classification techniques~\cite{Sultan2018,Mendels2018,Bonati2020,trizio2021enhanced}.
In one of such methods, called Deep Targeted Discriminant Analysis (DeepTDA)~\cite{trizio2021enhanced}, a neural network is optimized to obtain a low-dimensional representation ($s$) in which the probability distribution of the metastable states corresponds to a predefined mixture of Gaussians.
This is achieved by using a loss function that acts on the mean values and variances of the distributions of each of the $n_c$ classes (states) in the $s$ space:
\begin{equation}
\mathcal{L}_{\textit{DeepTDA}} = \sum_{c=1} ^{n_C} \left|\mu_c(s)-\mu_c^{target}\right|^2 + \left|\sigma_c(s)-\sigma_c^{target}\right|^2 \label{eqn:9}
\end{equation}
A question that follows naturally is how these CVs can be improved as new data outside the metastable states become available. It is indeed not easy to correctly assign labels to generic system configurations.
In our context, we want to learn the CVs using also the data collected from TPS trajectories. 
In a variant of the method called transition path informed (TPI-DeepTDA)~\cite{parrinello20233}, data belonging to the transition region were assigned to a third class, requiring them to be mapped onto an additional Gaussian located between states and with a broader distribution.
However, this could make the results highly dependent on the definition of metastable states.

A more general solution to the problem is based on a multi-task approach, where a single model is optimized on different data sets according to different criteria~\cite{parrinello20232}.
This allows us to learn the CV in a semi-supervised way: on the labeled data, it is optimized to distinguish between states, while on the data without labels, it is optimized in an unsupervised manner, maximizing the information content of the CV.

From a practical point of view, we follow the implementation of Ref.~\cite{parrinello20233}, in which a semi-supervised CV is achieved by combining Deep-TDA with an autoencoder.
An autoencoder is an artificial neural network consisting of two parts: an encoding function $E$ that transforms the input data into a (typically) lower-dimensional representation (here, the CV) and a decoding function $D$ that tries to reconstruct the initial data from this compressed representation:
\begin{equation}
\mathcal{L}_{AE} = \sum_{i=1} ^{n_{data}} |x_i-D \circ E(x_i)|^2\label{eqn:10}
\end{equation}
The loss function for the multi-task CV is given by the linear combination of the reconstruction loss (calculated on the dataset without labels) and the DeepTDA loss (on the labeled dataset) acting on the bottleneck $s$:
\begin{equation}
\mathcal{L}_{multitask} = \mathcal{L}_{AE}+\alpha\ \mathcal{L}_{DeepTDA}\label{eqn:11}
\end{equation}
where $\alpha$ is a parameter that gives the relative weight of the two losses.
This means the resulting CV is optimized to reconstruct the data as in a standard autoencoder and discriminate between states.
Therefore, we see this approach as regularizing the latent space learned from autoencoders.

In our case, the labeled dataset is constructed from short, unbiased simulations of the metastable states, while the configurations extracted from the transition path sampling trajectories compose the unlabeled one.
In this way, we can fulfill the requirement that CVs should distinguish metastable states from the reactive path and the transition state, without explicitly labeling the latter.
By doing so, fitting the shooting point density in this CV space will lead to an enhanced efficiency of TPS simulations. Furthermore, once the CV has been optimized, the free energy profile of the reactions can be effectively reconstructed using biased enhanced sampling methods such as Umbrella Sampling, Metad or OPES (see the Appendix in the supporting information for further details about free energy calculations). 

\subsection*{The MLCV-based RB-shooting workflow.}

Putting all the pieces together, we propose a new shooting algorithm for TPS simulations based on the concept of reactivity, measured as a function of a set of collective variables. The CVs are extracted from the simulation data using a multi-task objective function. The density of successful shooting points is fitted in the CV space and used to bias the selection of shooting points toward higher reactivity. This allows us to systematically improve TPS simulations' efficiency without prior knowledge of the process. At the end of this procedure, we obtain an unbiased set of reactive trajectories together with optimized reaction coordinates and, with minimal additional effort, also the free energy profiles. The complete workflow of our approach is given below:

\begin{enumerate}
    \itemsep0em 
    \item Run conventional MD runs in each metastable state to collect configurations for the supervised datasets.
    \item Run bootstrap two-way shooting TPS simulations to collect initial transition paths (unsupervised datasets) and shooting point structures.
    \item Train the MLCV with supervised and unsupervised datasets.
    \item Perform density estimations for both reactive and all shooting points ($\tilde{\rho}^{sp}_{reactive}$ and $\tilde{\rho}^{sp}_{total}$) in Eq.~\ref{eqn:7}.
    \item Run RB-shooting TPS using the biasing shooting point selection function in Eq.~\ref{eqn:8}, and collect the newly generated transition paths and shooting points.
    \item Go back to step 3 to update the MLCV and shooting point densities. When the shooting success rate is satisfying, proceed to step 7.
    \item Freeze the shooting point biasing selection function and run production simulations.
    \item Perform biased free energy calculations based on the final MLCVs.
\end{enumerate}

\subsection*{Computational Setup}

Here we describe the key details of the systems simulated and the parameters employed in the RB-shooting workflow, while further computational details can be found in the supplementary information. 

\if\ispreprint1
\paragraphtitle{Brownian Particle on 2-Dimensional Potential Energy Surface.}
\else
\textbf{Brownian Particle on 2-Dimensional PES.}
\fi 
The potential energy surface was defined as:
\begin{equation}
U(x,y) = B\big((x^2-1)^2 + (x-y)^2\big), \label{eqn:12}
\end{equation}
where $B$ is the barrier height.
Based on the PES, we defined the two configurational space regions with potential energy lower than $0.1\ B$ as the two metastable states (Fig.~\ref{fig:1}A).
To bootstrap the RB-shooting, we ran 100 ps unbiased simulations from both metastable states, in addition to a conventional two-way TPS simulation, including 400 MC moves.
From the harvested trajectories, we constructed the initial datasets that contain 10093 labeled conformations and 10000 unlabeled conformations.
Using this dataset, we performed RB-shooting TPS simulations using the semi-supervised multi-task CV \cite{parrinello20232}. The optimization of the MLCVs was performed using the \verb|mlcolvar|~\cite{parrinello20232} library. Furthermore, to test the effect of using different MLCVs, we also performed another set of simulations using the TPI-DeepTDA CV \cite{parrinello20233}.
Detailed network architectures are listed in the supporting information.
Both simulations undergo the two-phase workflow mentioned above.
During each biasing function optimization phase, 100 RB-shooting MC moves were made, then the MLCV and the shooting point densities were updated.
When updating MLCVs, the initial labeled dataset was used again, and the unlabeled dataset was constructed from the updated TPE with a maximum data point number of 10000.
After six optimization steps, the biasing function was frozen, and a 50000-step production TPS run was performed using the final selection function.
In all simulations, the density cutoff value in Eq.~\ref{eqn:8} was selected as 0.1.
Besides, the maximum path length was constrained to 30 ps.
To determine the efficiency improvement resulting from the RB-shooting, we carried out a 50000-step uniform two-way TPS simulation.
Besides, we also obtained 40000 transition paths from conventional MD runs starting from the metastable states, to assert the convergence of all TPS runs.

\if\ispreprint1
\paragraphtitle{Alanine Dipeptide.}
\else
\textbf{Alanine Dipeptide.}
\fi
We sampled the configurational transition of the alanine dipeptide in explicit solvent.
The metastable states were defined in the 2-dimensional space spanned by the backbone dihedral angles $\phi$ and $\psi$: the A state includes conformations that locate in the $-180^{\circ} < \phi < -75^{\circ}$ region, and the B state includes conformations that locate in the circle centered at $[60.2^{\circ}, -114.6^{\circ}]$ and with a radius of $20^{\circ}$ (see Fig.~\ref{fig:3}).
To bootstrap the RB-shooting, we ran a 100 ns unbiased simulation from the B states, in additional to a uniform two-way TPS run including 800 MC moves.
From the resulting trajectories, we constructed an initial dataset that contains 25000 labeled conformations and 25000 unlabeled conformations.
Using the dataset, we carried out RB-shooting TPS simulations using the multi-task CV \cite{parrinello20232}.
Detailed network architectures are listed in the supporting information.
The procedure was composed by three biasing function optimization iterations and a 20000-step production TPS run that uses the optimized selection function.
During each optimization step, 200 MC moves were made.
When updating MLCVs, the initial labeled dataset was used again, and the unlabeled dataset was constructed from the updated TPE with a maximum data point number of 25000.

To measure the efficiency improvement brought by the RB-shooting scheme, we performed another uniform two-way TPS simulation of 20000 steps.
Besides, a one-way TPS simulation of 50000 steps was performed with the OpenPathSampling \cite{bolhuis20191,bolhuis20192} package to assess the convergence of all other TPS runs.
In all TPS runs, the maximum path length was constrained to 5 ps.
We also demonstrated that using the MLCV trained during the RB-shooting scheme, free energy changes of the conformation transition process can be accurately evaluated.
To this end, we carried out a 30 ns multi-task CV-based OPES simulation. The reference free energy was obtained from a 250 ns OPES simulation, biasing the$\phi$ and $\psi$ variables.
We reweighted the calculated free energies alone the torsion variable $\phi$, and the free energy difference between the $-180^{\circ} < \phi < 10^{\circ}$ region and the $10^{\circ} < \phi < 126^{\circ}$ region was calculated.

\if\ispreprint1
\paragraphtitle{Hydrolysis of Acetyl Chloride.}
\else
\textbf{Hydrolysis of Acetyl Chloride.}
\fi
The hydrolysis of acetyl chloride in bulk water is a classical bimolecular nucleophilic substitution ($\mathrm{S_N2}$) reaction.
We use this reaction to examine the capability of our method to investigate chemical processes.
However, pure \textit{ab initio} and QM/MM potentials are computationally unaffordable for large-scale TPS simulations.
To overcome these limitations, we constructed a machine learning-based neuroevolution potential (NEP) \cite{fan2021,fan2022} to drive the sampling. The workflow used to train the ML potential is presented in the results, and the detailed setup can be found in the supporting information.

In the RB-shooting run, the two metastable states were defined by the water coordination number of the carbon atom in the acyl chloride group ($\mathrm{C_{C-O}}$), and the distance between this carbon atom and the chloride atom ($\mathrm{d_{C-Cl}}$): the reactant conformation state includes all conformations that satisfy $\mathrm{C_{C-O}} < 0.1$ and $\mathrm{d_{C-Cl}} < 0.185\ \mathrm{nm}$, the product conformation state includes all conformations that satisfy $\mathrm{C_{C-O}} > 0.7$ and $\mathrm{d_{C-Cl}} < 3.75\ \mathrm{nm}$ (see Fig.~\ref{fig:4}).
With these state definitions, we constructed the initial dataset from the \textit{ab initio} SMD trajectory and a 700-step bootstrapping two-way TPS run.
The resulting dataset contains 7602 labeled conformations and 10000 unlabeled conformations.
After obtaining the dataset, the RB-shooting TPS simulation was carried out using the multi-task CV \cite{parrinello20232}.
Detailed network architectures are listed in the supporting information.
Since the above RB-shooting TPS simulation aimed to build a reliable MLCV to support the subsequent MetaD-based active learning, it was only composed of two biasing function optimization iterations.
During each optimization iteration, 200 MC moves were made.
When updating MLCVs, the initial labeled dataset was used, while the unlabeled one was constructed from the updated TPE with a maximum number of data points equal to 10000.

After fitting the final machine learning potential, we performed the production RB-shooting TPS run.
This TPS run has the same parameters as the previous one but contains 20000 extra MC moves using the optimized basing function.
To determine the efficiency improvement brought by RB-shooting, we carried out another 20000-step uniform two-way TPS simulation.
Furthermore, a 30000-step one-way TPS simulation was performed to assert the convergence of the RB-shooting TPS runs.
The maximum path length was constrained to 5 ps in all TPS runs.
Finally, we calculated the free energy profile of this reaction by the mean of Umbrella Sampling \cite{torrie1977}, using the MLCV optimized from the production RB-shooting TPS run as collective variable.

\section{Results And Discussions}

\subsubsection{Brownian Particle on 2-Dimensional Potential Energy Surface.}

\if\ispreprint1
    \begin{figure*}[htb!]\centering
    \includegraphics[width=\linewidth]{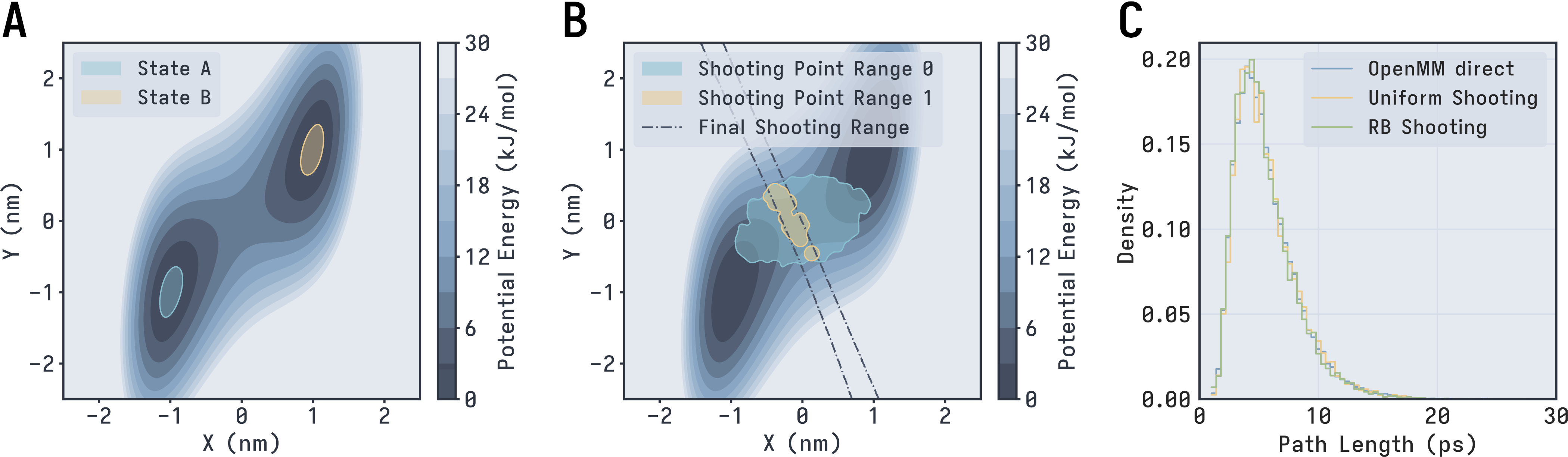}
    \caption{(A) Potential energy surface and state definitions of the 2D Brownian particle system. (B) Shooting ranges evolution during the biasing function optimization. The final shooting range is represented by the region between the two dashed lines. The MLCV used here is the multi-task CV. (C) Path length distribution resulting from each sampling method.}
    \label{fig:1}
    \end{figure*}
\else
    \begin{figure}\centering
        \includegraphics[width=\linewidth]{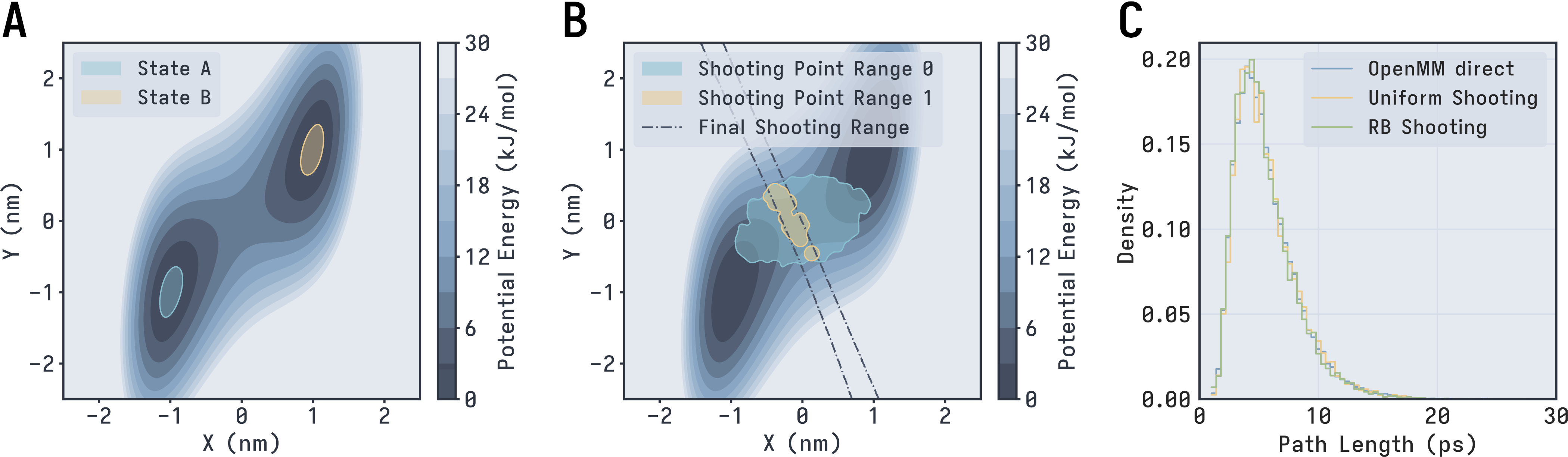}
        \caption{(A) Potential energy surface and state definitions of the 2D Brownian particle system. (B) Shooting ranges evolution during the biasing function optimization. The final shooting range is represented by the region between the two dashed lines. The MLCV invoked here is the multi-task CV. (C) Path length distribution resulting from each sampling method.}
        \label{fig:1}
    \end{figure}
\fi
\if\ispreprint1
    \begin{figure*}\centering
    \includegraphics{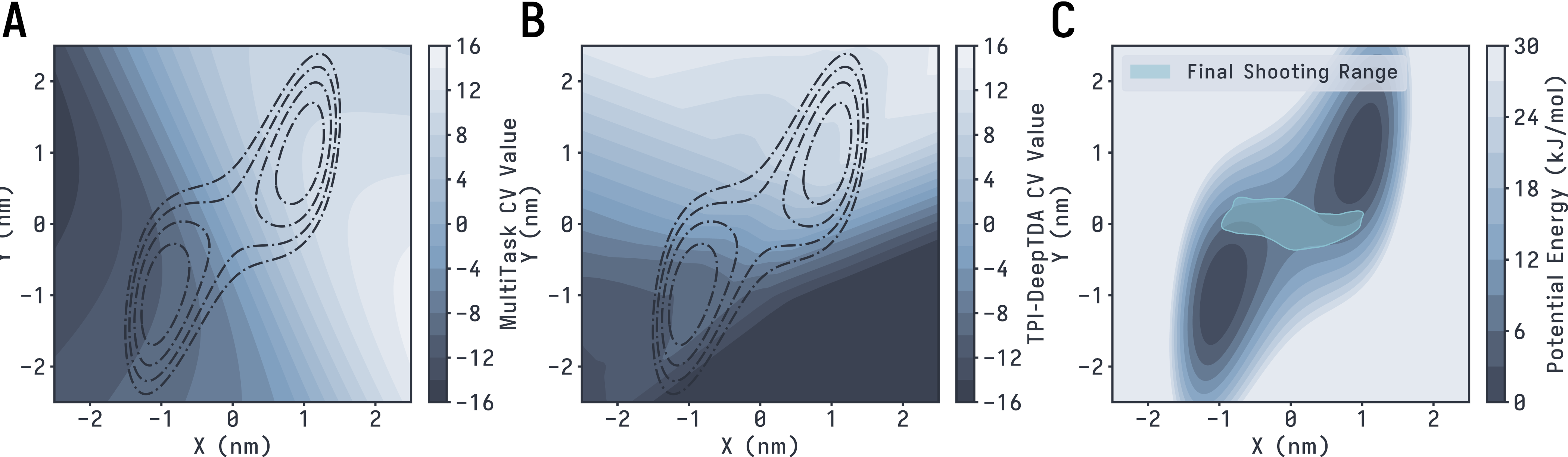}
    \caption{(A) Values of the multitask CV in the configuration space of the 2D PES system. (B) Values of the TPI-DeepTDA CV in the configuration space of the 2D PES system. (C) Final shooting range produced in the TPI-DeepTDA CV-based RB-shooting TPS simulation of the 2D PES system.}
    \label{fig:2}
    \end{figure*}
\else
    \begin{figure}\centering
    \includegraphics[width=\linewidth]{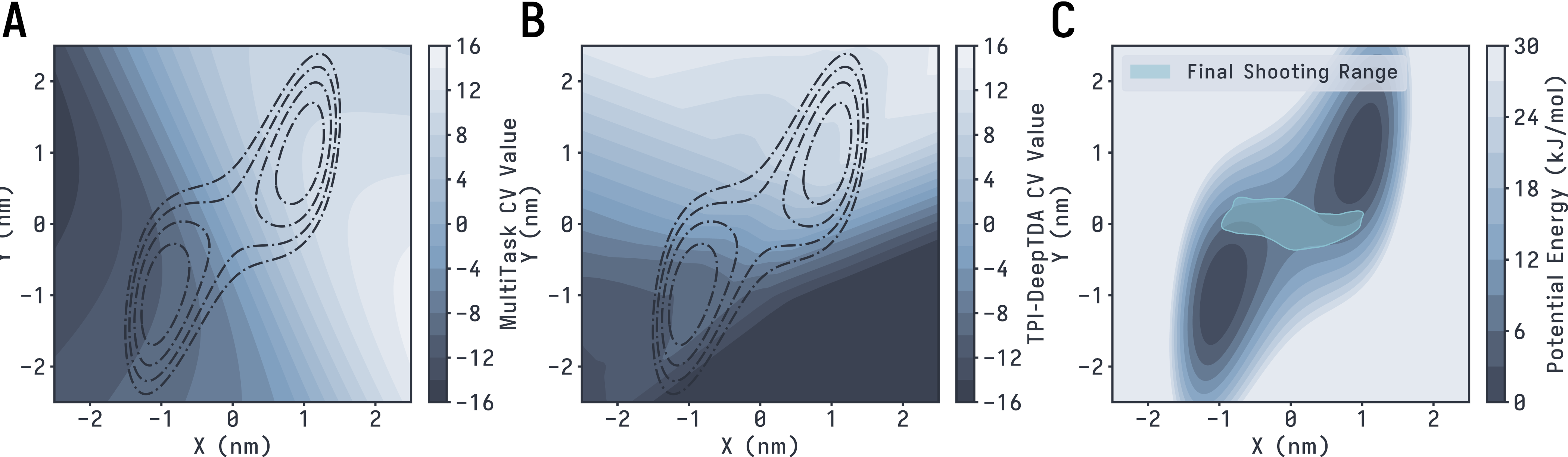}
    \caption{(A) Values of the multitask CV in the configuration space of the 2D PES system. (B) Values of the TPI-DeepTDA CV in the configuration space of the 2D PES system. (C) Final shooting range produced in the TPI-DeepTDA CV-based RB-shooting TPS simulation of the 2D PES system.}
    \label{fig:2}
    \end{figure}
\fi
We start by analyzing the behavior of our workflow in the 2D toy-model system.
The following results were obtained using multi-task CV-based runs; different CV learning schemes will be compared later.
First, we examine the convergence of the shooting ranges during the biasing function optimizations.
In Fig.~\Ref{fig:1}B, we plotted the ranges of successful shooting points (shooting points that lead to reactive trajectories):
the initial range before the optimization is shown in blue, and the range after the first optimization iteration is shown in yellow.
Clearly, after one single iteration of MLCV fitting and shooting point density estimation, the shooting range was concentrated at the transition region.
Furthermore, the shooting range after the first optimization iteration is similar to the final shooting range after six iterations of biasing function optimization (regions between the two dashed lines in Fig.~\Ref{fig:1}B).
This means, at least for simple processes, our shooting method converges quickly.

Then, we checked the success rate of our shooting move.
In the multi-task CV-based RB-shooting simulation, 23989 reactive trajectories were generated from the 50000 production MC moves, which leads to a success rate of 48.0\%.
To compare with, in the uniform two-way TPS simulation, 11545 reactive trajectories were generated from the 50000 production MC moves, which leads to a success rate of 23.1\%.
Hence, the RB-shooting TPS showed an over 2-fold improvement in shooting efficiency.
Besides efficiency, checking if our shooting scheme can generate correct path ensembles is critical.
To this end, we plotted the distribution of the reaction path lengths in Fig.~\Ref{fig:1}C, which shows how uniform two-way shooting and RB-shooting-based TPS simulations could generate the same path length distributions as brute-force MD simulations.

We also analyzed the effects of using different MLCV fitting strategies. In particular, we compared the multi-task CV approach with the TPI-DeepTDA CV~\cite{parrinello20233}. As discussed in the methods, both approaches allow us to combine data from the equilibrium states with data from the transition state region. 
As shown in Table~\Ref{tbl:1}, the TPI-DeepTDA CV-based RB-shooting also improves shooting success rate, although not as significant as that shown in the multi-task CV-based RB-shooting TPS run.
To elucidate the causes, we plotted the CV values in the 2D configuration space in Fig.~\ref{fig:2}A and Fig.~\ref{fig:2}B.
It is easy to find out that the multi-task CV presented a better resolution in the transition region.
The TPI-DeepTDA CV, otherwise, sightly mixed the transition region with the metastable regions, which could also be told from the final shooting range (Fig.~\ref{fig:2}C).
This phenomenon may be caused by TPS trajectories containing not only configurations of the transition region, but also structures more similar to metastable ones (which also have a higher density in configurational space). One could alleviate this problem by filtering the trajectories to take only configurations close to the transition state, but this is not a straightforward procedure. 
On the other hand, the multi-task CV does not make any assumption on the distribution of the training set, but only tries to maximize the structural information content of the TS configurations in the lower-dimensional space.
As a result, the multi-task CV can better distinguish each TPE configuration, which is a crucial feature to fit the shooting point densities and achieve a higher success rate.
Based on this evidence, we believe that the multi-task CV is a better choice for RB-shooting TPS simulations, and we only used this MLCV fitting scheme for the other two systems.

\if\ispreprint1
    \begin{table}[b!]
  \caption{Success rates of different shooting methods per the different systems: the 2D toy model, Alanine Dipeptide (Ala2) and the Hydrolysis of Acetyl Chloride (S$_\text{N}$2). The improvement (ratio) is given with respect to the reference two-way shooting.}
  \label{tbl:1}
  \begin{tabular}{llll}
    \hline
    System & Shooting Method & Success Rate & Ratio\\
    \hline
    2D PES & Uniform Two-way & 23.1\% & - \\
    2D PES & RB-shooting (multitask CV) & 48.0\% & 2.1× \\
    2D PES & RB-shooting (TPI-DeepTDA CV) & 36.2\% & 1.6× \\
    \hline Ala2 & Uniform Two-way  & 23.7\% & - \\
    Ala2 & RB-shooting (multitask CV) & 42.8\% & 1.8× \\
    \hline S$_\text{N}$2  & Uniform Two-way  & 5.8\% & - \\
    S$_\text{N}$2 & RB-shooting (multitask CV) & 46.4\% & 8.0× \\
    \hline
  \end{tabular}
    \end{table}
\else
  \begin{table}
  \caption{Success rates of different shooting methods per the different systems: the 2D toy model, Alanine Dipeptide (Ala2) and the Hydrolysis of Acetyl Chloride (S$_\text{N}$2). The improvement (ratio) is given with respect to the reference two-way shooting.}
  \label{tbl:1}
  \begin{tabular}{llll}
    \hline
    System & Shooting Method & Success Rate & Ratio\\
    \hline
    2D PES & Uniform Two-way & 23.1\% & - \\
    2D PES & RB-shooting (multitask CV) & 48.0\% & 2.1× \\
    2D PES & RB-shooting (TPI-DeepTDA CV) & 36.2\% & 1.6× \\
    \hline Ala2 & Uniform Two-way  & 23.7\% & - \\
    Ala2 & RB-shooting (multitask CV) & 42.8\% & 1.8× \\
    \hline S$_\text{N}$2  & Uniform Two-way  & 5.8\% & - \\
    S$_\text{N}$2 & RB-shooting (multitask CV) & 46.4\% & 8.0× \\
    \hline
  \end{tabular}
    \end{table}
\fi

\if\ispreprint1
    \begin{figure*}\centering
    \includegraphics{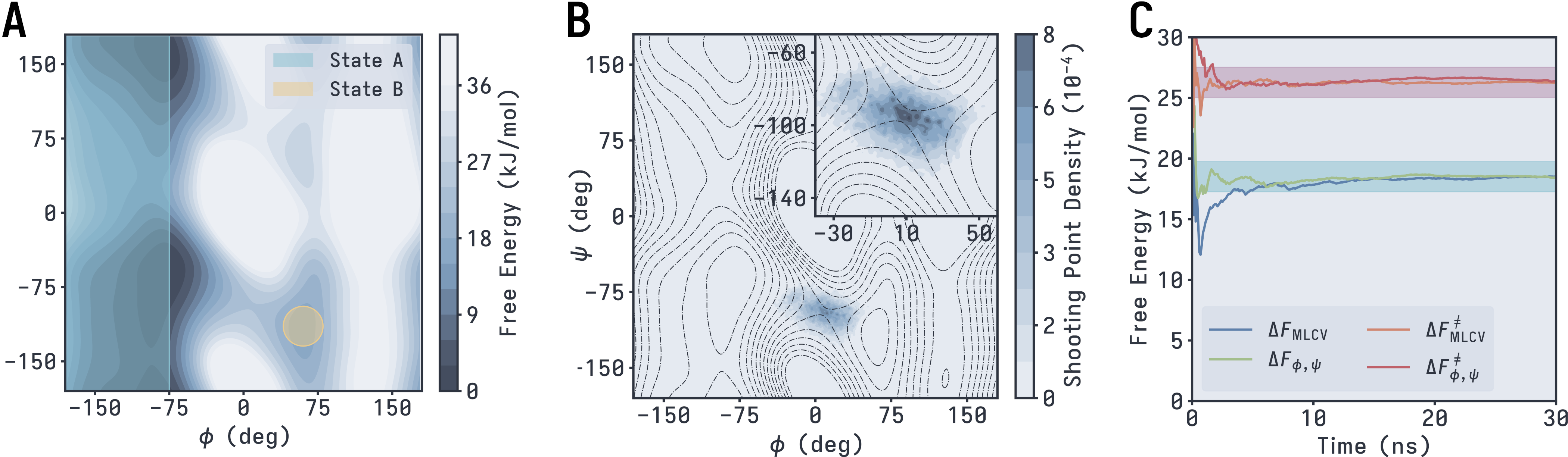}
    \caption{(A) Free energy surface and state definitions of the alanine dipeptide system. (B) Shooting point density generated during the production run. (C) Free energy difference between the two torsional space regions, and the free energy barrier during the multi-task CV-based OPES free energy calculation of the alanine dipeptide system. The reference value was obtained from the 250 ns OPES simulation using torsions as CV, and the $\pm 0.5 k_{\mathrm{B}}T$ range is represented by the colored regions.}
    \label{fig:3}
    \end{figure*}
\else
    \begin{figure}\centering
    \includegraphics[width=\linewidth]{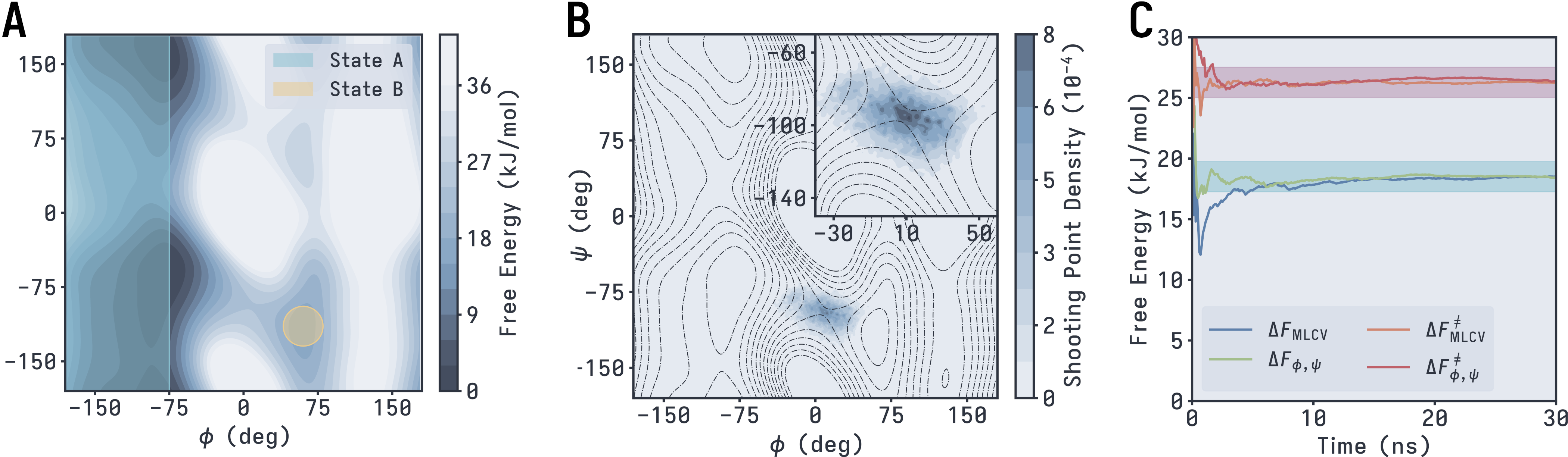}
    \caption{(A) Free energy surface and state definitions of the alanine dipeptide system. (B) Shooting point density generated during the production run. (C) Free energy difference between the two torsional space regions, and the free energy barrier during the multi-task CV-based OPES free energy calculation of the alanine dipeptide system. The reference value was obtained from the 250 ns OPES simulation using torsions as CV, and the $\pm 0.5 k_{\mathrm{B}}T$ range is represented by the colored regions.}
    \label{fig:3}
    \end{figure}
\fi

\subsubsection{Alanine Dipeptide.}

Alanine dipeptide is one of the standard systems for benchmarking sampling algorithms.
In this system, the rare dominant transitions happen in the space spanned by the two Ramachandran torsional angles $\phi$ and $\psi$.
In Fig.~\ref{fig:3}A we show the free energy surface on the $\phi$ and $\psi$ variables and the definition of the metastable states.

First, we look into the simulation efficiency and the correctness of the sampled path ensemble.
In the uniform two-way TPS simulation, which is our baseline here, 4745 reactive trajectories were generated from the 20000 production MC moves, which leads to a success rate of 23.7\%. In the RB-shooting simulations, 8562 reactive trajectories were generated from the 20000 production MC moves, which leads to a success rate of 42.8\%.
The RB-shooting again significantly improves the sampling efficiency, while at the same time also decreasing the autocorrelation of the generated paths with respect to uniform two-way shooting (Fig. S2B).
In Fig. S1A, we show the distribution of reaction path lengths sampled using different protocols.
The good agreements between the RB-shooting TPS simulation and the reference one-way TPS simulation confirm the correctness of our results.

Then, we show the shooting range produced by RB-shooting.
From Fig.~\ref{fig:3}B, we find that the shooting range surrounds the high energy transition region as expected.
Since our biasing function is essentially a set of truncated Gaussian functions, shooting points could be selected from the entire transition region, but with a higher chance at the center of this region.
This behavior ensures a high shooting success rate and an acceptable decorrelating speed.

Finally, we check the convergence speed of the MLCV-based free energy calculation.
As shown in Fig.~\ref{fig:3}C, the free energy difference between the two states converged within $\pm 0.5 k_{\mathrm{B}}T$ of the reference value in less than 5 ns (blue line).
The convergence speed is similar to the OPES simulation carried out in the 2D torsional space (green line in Fig.~\ref{fig:3}C).
The barrier height converges faster than the metastable states' free energy difference. For the MLCV-based OPES run, this value converged to the reference within 2 ns, which is even a little faster than the dihedral-based runs.
These results suggest that the MLCV could precisely capture critical features of the conformational transition and is thus suitable for accurate and efficient free energy calculation.

\subsubsection{Hydrolysis of Acetyl Chloride.}

The hydrolysis of acetyl chloride is one of the most classical $\mathrm{S_N2}$ reactions. Nevertheless, we want to use it as an example of a realistic application, showing how our workflow can be successfully employed without prior knowledge of the reaction mechanism. To study the reaction with high accuracy, we first integrated our workflow into an active learning scheme to iteratively collect configurations along the transition pathways and build a reliable reactive potential. To construct the ML potential without assuming the reaction mechanism, we adopted the following workflow:
\begin{enumerate}
    \itemsep0em 
    \item Obtain an initial reactive trajectory from an  \textit{ab initio} steered MD (SMD) simulation that stretches the distance between the chloride atom and the carbon atom that bonds with it.
    \item Uniformly select structures from the reactive trajectory and train an initial machine learning potential from the selected structures.
    \item Perform SMD-based active learning \cite{parrinello2022} to obtain a more robust machine learning potential.
    \item Run RB-shooting TPS simulations using the trained potential and save the resulting MLCV.
    \item Perform MetaD-based active learning using the machine learning potential from step 3 and the MLCV from step 4.
    \item Select transition state structures from the MetaD trajectories in step 5 using the optimized shooting range in step 4, and perform \textit{ab initio} calculations on these structures.
    \item Train the final machine learning potential with the training set obtained in step 5 and the selected structures from step 6.
\end{enumerate}
Enhanced sampling techniques are crucial for obtaining a reliable reactive potential. SMD simulations (step 3) provide the first guess of the reactive pathway, while metadynamics simulations (step 5) allow us to obtain a more thorough sampling. Finally, adding TS configurations ensures uniform accuracy along the whole transition pathways, as shown in Ref.~\cite{parrinello2022}.  When this procedure is completed, we use the ML-based potential to run production RB-shooting simulations. Although the hydrolysis of acetyl chloride is a simple and well-known process, to our knowledge, this is one of the first research works that use  TPS simulations assisted by ML potentials to study the mechanisms of chemical reactions with an \textit{ab initio} quality. 

\if\ispreprint1
    \begin{figure*}\centering
    \includegraphics{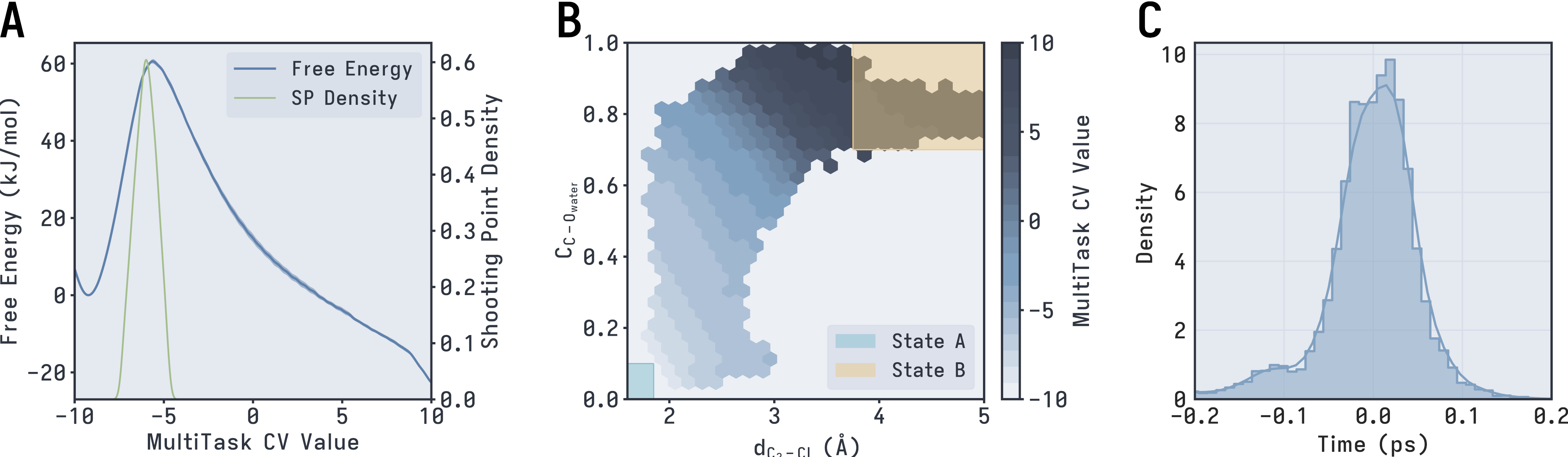}
    \caption{(A) Free energy profile of the hydrolysis of acetyl chloride, and the shooting point density generated by RB-shooting. (B) Multi-task CV values are projected on the two dominant descriptors and the state definitions. (C) Time gap between the forming of the carbon-oxygen bond and the breaking of the chloride-carbon bond.}
    \label{fig:4}
    \end{figure*}
\else
    \begin{figure}\centering
    \includegraphics[width=\linewidth]{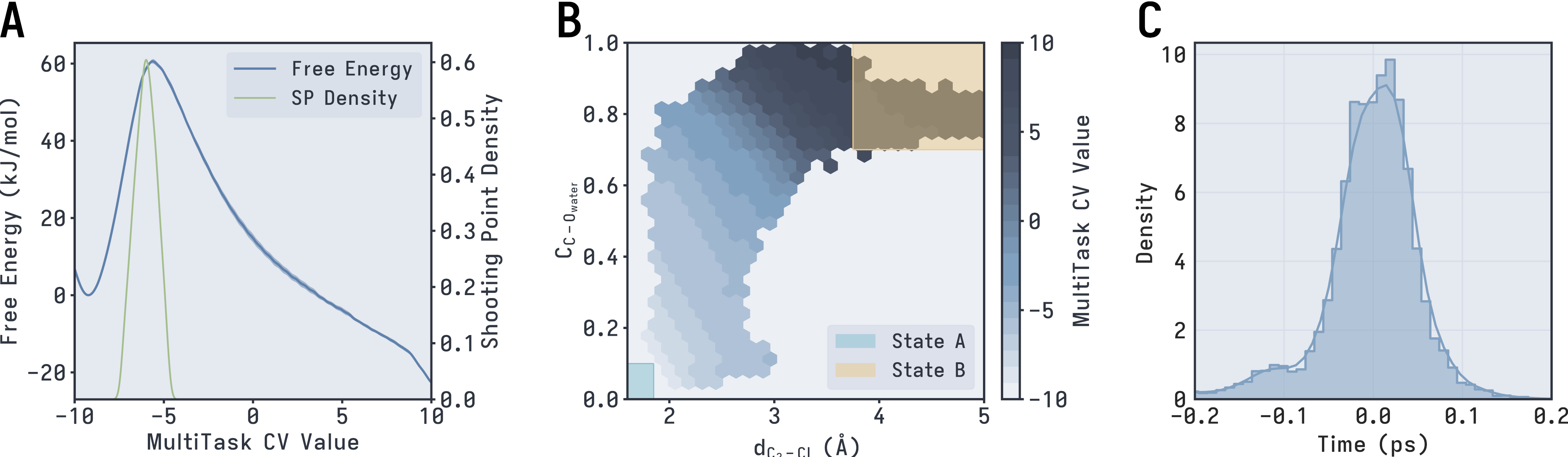}
    \caption{(A) Free energy profile of the hydrolysis of acetyl chloride, and the shooting point density generated by RB-shooting. (B) Multi-task CV values are projected on the two dominant descriptors and the state definitions. (C) Time gap between the forming of the carbon-oxygen bond and the breaking of the chloride-carbon bond.}
    \label{fig:4}
    \end{figure}
\fi

First, we look into the simulation efficiency and the correctness of the sampled path ensemble. In the baseline uniform two-way TPS simulation, 1158 reactive trajectories were generated from the 20000 production MC moves, which leads to a success rate of 6.8\%. In the RB-shooting simulation, 9287 reactive trajectories were generated from the 20000 production MC moves, which leads to a success rate of 46.4\%. Thus, the RB-shooting algorithm brings an 8-fold improvement to the shooting efficiency, and the generated pathways are significantly less correlated than those generated by the uniform two-way shooting (Fig. S2C).
The reason for such a large improvement may be as follows: the reaction channel of this process is rather narrow; thus shooting points near the metastable states have a meager chance of entering the channel.
In such a situation, the RB-shooting scheme, which only selects shooting points closer to the transition region, can result in greater efficiency.
Besides, from Fig. S1B, we could also assert the correctness of the TPE sampled by RB-shooing.

As shown in Fig.~\ref{fig:4}A, the free energy barrier calculated along the MLCV is about 60 kJ/mol, which is very close to the experimental value of about 65 kJ/mol \cite{ruff2010}.
The tiny sampling error (shown as the blue shadow around the free energy curve in Fig.~\ref{fig:4}A) also suggests the robustness of both the MLCV fitting scheme and the entire free energy calculation.
In the same panel, we plotted the shooting point density produced by RB-shooting.
As expected, the density is centered at the transition region (the free energy maximum).
These phenomena indicate the excellent ability of the MLCV to distinguish different conformation states.
Furthermore, we showed the CV values projected on the 2D space spanned by $\mathrm{C_{C-O}}$ and $\mathrm{d_{C-Cl}}$ (Fig.~\ref{fig:4}B).
The distinction of CV values in different configuration regions clearly illustrates the great resolution of the MLCV.

Finally, we investigate the ability of our workflow to identify the correct reaction mechanism. Being a $\mathrm{S_N2}$ reaction, we know that a water molecule binds to the carbon atom of the chloride acyl group and the chloride-carbon bond is broken at the same time, but in general, we won't have such information in advance. 
To stress-test our workflow, we assumed another type of mechanism, specifically an unimolecular nucleophilic substitution $\mathrm{S_N1}$, to test if the RB-shooting scheme could recover the actual mechanism. TPS simulations were initiated from an \textit{ab initio} SMD simulation in which the chloride-carbon bond broke first and the oxygen-carbon bond formed after 5.2 ps. To analyze the reaction mechanism identified by the final TPS runs, we first obtained the ensemble-averaged transition state geometry from structures within the shooting range (Fig. S3C) and used it as the criterion for judging the forming and breaking of relative chemical bonds.
In Fig.~\ref{fig:4}C, we plotted the distribution of the time gap between the forming of the carbon-oxygen bond and the breaking of the chloride-carbon bond, as obtained from the TPE.
The distribution is centered at 0, which implies that the two processes happen simultaneously, as it should be in a $\mathrm{S_N2}$ reaction.
These results demonstrate the great potential of RB-shooting TPS in investigating chemical reactions: even starting from a wrong initial guess, our workflow could recover the true mechanism with high efficiency.

\section{Conclusions}

This work introduced an efficient shooting move for transition path sampling.
The proposed workflow relies on the density fitting of shooting points in low-dimensional reaction coordinate spaces.
To reduce the dependency on prior knowledge about the sampled processes, we leveraged recent advances in the machine learning-based collective variables, which allowed us to learn such variables directly from the simulation data.
Besides, using the fitted MLCVs, free energy changes of the interested processes can be accurately evaluated using biased enhanced sampling protocols.
As a result, our workflow can be seen as a synergistic combination of the two families of enhanced sampling methods, in the same vein as a recent combination of transition path sampling and metadynamics~\cite{dellago2023}.

We tested our workflow with three different processes: the Brownian motion of a particle on a 2D PES, alanine dipeptide conformational transitions, and acetyl chloride hydrolysis in bulk water.
For each process, we obtained the correct transition path ensemble compared with conventional TPS variants but with much higher efficiency.
We also achieved fast and precise free energy estimations using the CVs with biased enhanced sampling methods.
In the study of the hydrolysis of acetyl chloride, we combined our workflow with an active learning strategy for fitting machine learning potentials to DFT reference calculations. This allowed us to obtain an \textit{ab initio}-quality description of the reaction, resulting in a free energy barrier very similar to the experimental one.

The above results demonstrate the usefulness of our workflow. Furthermore, they show that if the collective variables can effectively distinguish the transition path configurations, it is possible to select shooting ranges from regions of high reactivity to improve the shooting success rate.
Regarding future developments, it is important to note that CVs are not limited to being one-dimensional committor-like functions but can also be multi-dimensional. Hence, this provides a natural framework to extend the RB shooting method to sample multi-state transition paths. Moreover, integrating recent methodologies developed to sample the transition region through biased enhanced sampling \cite{parrinello2024} into this approach might result in further improvement, making it more robust even in cases where it is difficult to obtain an initial guess from standard TPS. 
Therefore, we believe this approach could play a significant role in extending molecular dynamics simulations and can be applied to realistic systems to shed light on both the reaction mechanisms and reconstruct the free energy profiles.

\section*{Acknowledgements}
This work received support from the National Natural Science Foundation of China (22220102001 and 92370130).
L.B. acknowledges funding from the Federal Ministry of Education and Research, Germany, under the TransHyDE research network AmmoRef (support code: 03HY203A).
The authors thank Timothée Devergne and Enrico Trizio for providing feedback on the manuscript and Dong Yang for many valuable discussions in the field of \textit{ab initio} calculations.

\bibliography{main}

\end{document}